\documentstyle[aps,twocolumn]{revtex}

\draft

\begin{document}
\title{Size effects on excitons in nano-rings}
\author{Hui Hu$^a$, Dai-Jun Li$^a$, Jia-Lin Zhu$^{a,b}$, Jia-Jiong Xiong$^a$}
\address{$^a$ Department of Physics, Tsinghua University, Beijing 100084, P. R. China%
\\
$^b$ Center for Advanced Study, Tsinghua University, Beijing 100084, P. R.\\
China}
\date{\today}
\maketitle

\begin{abstract}
The size effects of an exciton in a nano-ring are investigated theoretically
by using an effective-mass Hamiltonian which can be separated into terms in
center-of-mass and relative coordinates. The binding energy and oscillator
strength of the ground state are calculated for two different ring radii as
a function of the ring width. The resulting linear optical susceptibility of
the low-lying exciton states is also discussed. \newline
{PACS numberes: 73.20.Dx, 71.35.-y, 78.66.Fd} \newline
\end{abstract}

Recent progress in nanofabrication techniques has made it possible to
construct self-assembled InGaAs nano-rings \cite
{warburton,lorke00,pettersson,lorke98,lorke99}. Quite different from the
previously fabricated sub-micron GaAs quantum rings \cite{mailly}, those
nano-rings now achieved are so small (with characteristic inner/outer radius
of 20/100 nm and 2-3 nm in height), that the electrons and holes can
propagate coherently (non-diffusively) throughout the ring. In view of this,
nano-rings can be viewed as a promising candidate for application in
microelectronics as well as the conventional quantum dots. Moreover, the
additional non-simply-connected geometry of nano-rings is of inherent
interest at the moment \cite{wendler95,chaplik,halonen96,romer,song00}.

While the conventional quantum dots have been investigated theoretically and
experimentally in depth \cite{que,halonen92,song,uozumi,wojs}, nano-rings
with strong quantum effects have only been recently treated \cite
{halonen96,song00,chak,gudm,wendler96,emperador,koskinen,hu,borrmann}. In
particular, the theoretical results related to the quantum confinement
effects on {\em exciton} states in nano-rings are very rare. Only recently
did Song and Ulloa report numerical calculations of the binding energy and
electron-hole separation of the exciton in an external magnetic field \cite
{song00}. They claim that the excitons in nano-rings behave to a great
extent as those in quantum dots of similar dimensions.

In this paper, we would like to investigate the size effects on excitons of
nano-rings by introducing a simplified confining potential, which is
applicable to the {\em realistic} self-assembled semiconducting InGaAs
nano-rings achieved to date\cite{lorke00,lorke98,lorke99}. To explore the
role of different confinements, we express the model Hamiltonian in terms of
the center-of-mass and relative coordinates and calculate binding energies,
oscillator strengths and their dependence on the width of nano-rings, as
well as the linear optical susceptibility which would be measurable in
photoluminescence experiments, for example.

Our model is a two-dimensional exciton in a nano-ring, simulating recent
experimental nano-ring structures \cite{lorke00,lorke98,lorke99}. The
nano-ring is described by an electron-hole pair ($i=e,h$) with an effective
band edge mass $m_i^{*}$ moving in a x-y plane, and a ring-like confining
potential is introduced as $U\left( {\bf \vec{r}}_i\right) =\frac 1{2R_0^2}%
m_i^{*}\omega _i^2\left( {\bf \vec{r}}_i{}^2-R_0^2\right) ^2$, where $R_0$
is the radius of the ring and $\omega _i$ is the characteristic frequency of
the radial confinement, giving a characteristic ring width $W\approx 2\sqrt{%
\frac \hbar {2m_i^{*}\omega _i}}$ for each particle. The resulting model
Hamiltonian is thus given by: 
\begin{equation}
{\cal H}=\sum\limits_{i=e,h}\left[ \frac{{\bf \vec{p}}_i^2}{2m_i^{*}}%
+U\left( {\bf \vec{r}}_i\right) \right] -\frac{e^2}{4\pi \varepsilon
_0\varepsilon _r\left| {\bf \vec{r}}_e-{\bf \vec{r}}_h\right| },
\end{equation}
where ${\bf \vec{r}}_i=\left( x_i,y_i\right) $ and ${\bf \vec{p}}_i=-i\hbar 
{\bf \vec{\nabla}}_i$ denote the position vector and momentum operator, $%
\varepsilon _0$ is the vacuum permittivity, and $\varepsilon _r$ is the
static dielectric constant of the host semiconductor. It should be pointed
out that the present ring-like confining potential can be rewritten as $%
U\left( {\bf \vec{r}}_i\right) =\frac 12m_i^{*}\omega _i^2\left(
r_i{}-R_0\right) ^2\frac{\left( r_i{}+R_0\right) ^2}{R_0^2}$. If one
replaces the operator $r_i$ in factor $\frac{\left( r_i{}+R_0\right) ^2}{%
R_0^2}$ by its mean value $\left\langle r_i\right\rangle =R_0$, the
confining potential returns to the widely used parabolic form \cite
{lorke00,song00,emperador,hu}. On the other hand, for narrow rings (with
steep confinement) our confining potential gives a more realistic
description than does the parabolic form. In the latter, as pointed by Song
and Ulloa \cite{song00}, the associated wavefunctions fail in a real system
because the increased confinement may push the levels into the anharmonic
part of the potential and even produce deconfinement of carriers. Figs. (1a)
and (1b) display the shape of the ring potential with two different radii: $%
10$ and $30$ {\rm nm}. The solid and dashed lines correspond to the ring
width $W=16.7$ and $9.6$ {\rm nm}, respectively.

In terms of the relative coordinate ${\bf \vec{r}}={\bf \vec{r}}_e-{\bf \vec{%
r}}_h$ and center-of-mass coordinate ${\bf \vec{R}}=\frac{m_e^{*}{\bf \vec{r}%
}_e+m_h^{*}{\bf \vec{r}}_h}{m_e^{*}+m_h^{*}}$, the model Hamiltonian is
divided into 
\begin{eqnarray}
H &=&{\cal H}_{cm}\left( {\bf \vec{R}}\right) +{\cal H}_{rel}\left( {\bf 
\vec{r}}\right) +{\cal H}_{mix}\left( {\bf \vec{R},\vec{r}}\right) , 
\nonumber \\
H_{cm} &=&\frac{{\bf \vec{P}}_{cm}^2}{2M}+\frac{M\omega _{cm}^2}{2R_0^2}%
\left( {\bf \vec{R}}{}^2-R_0^2\right) ^2,  \nonumber \\
H_{rel} &=&\frac{{\bf \vec{p}}_{rel}^2}{2\mu }+\frac \mu 2\frac{\left(
m_h^{*3}\omega _e^2+m_e^{*3}\omega _h^2\right) }{M^3R_0^2}r^4-\mu \omega
_{rel}^2{}r^2  \nonumber \\
&&-\frac{e^2}{4\pi \varepsilon _0\varepsilon _rr},  \nonumber \\
H_{mix} &=&-2\mu \left( \omega _e^2-\omega _h^2\right) \left( {\bf \vec{R}%
\cdot \vec{r}-}\frac{{\bf \vec{R}}^3{\bf \cdot \vec{r}}}{R_0^2}\right) 
\nonumber \\
&&+\frac{\mu \omega _{rel}^2}{R_0^2}\left[ R^2r^2+2\left( {\bf \vec{R}\cdot 
\vec{r}}\right) ^2\right]  \nonumber \\
&&+2\mu \frac{\left( m_h^{*2}\omega _e^2-m_e^{*2}\omega _h^2\right) }{%
M^2R_0^2}{\bf \vec{R}\cdot \vec{r}}^3,
\end{eqnarray}
where $\mu =\frac{m_e^{*}m_h^{*}}M$ is the electron-hole reduced mass and $%
M=m_e^{*}+m_h^{*}$ is the total mass. We have also introduced a
center-of-mass frequency $\omega _{cm}=\sqrt{\frac{\left( m_e^{*}\omega
_e^2+m_h^{*}\omega _h^2\right) }M}$ and a relative frequency $\omega _{rel}=%
\sqrt{\frac{m_h^{*}\omega _e^2+m_e^{*}\omega _h^2}M}$.

The main purpose in the change of variable above is to use the solutions of $%
H_{cm}$ and $H_{rel}$ as a basis for solving the full Hamiltonian. Those
solutions, {\em i.e.}, labeled by $\psi _\lambda ^{cm}\left( {\bf \vec{R}}%
\right) $ and $\psi _{\lambda ^{\prime }}^{rel}\left( {\bf \vec{r}}\right) $%
, can be solved by the series expansion method \cite{zhu90,zhu97}. Here, $%
\lambda =\left\{ n_{cm},l_{cm}\right\} $ and $\lambda ^{\prime }=\left\{
n_{rel},l_{rel}\right\} $ represent the quantum number pair of the radial
quantum number $n$ and orbital angular-momentum quantum number $l$. Another
advantage coming from center-of-mass and relative separation is that we can
include the negative Coulomb interaction $-\frac{e^2}{4\pi \varepsilon
_0\varepsilon _rr}$ in $H_{rel}$, thus avoiding the well-known
poor-convergence of the parabolic basis when the characteristic system scale
is beyond the effective Bohr radius \cite{song00,song95}. We now search for
the wave functions of the exciton in the form 
\begin{equation}
\Psi =\sum\limits_{\lambda ,\lambda ^{\prime }}A_{\lambda ,\lambda ^{\prime
}}\psi _\lambda ^{cm}\left( {\bf \vec{R}}\right) \psi _{\lambda ^{\prime
}}^{rel}\left( {\bf \vec{r}}\right) .
\end{equation}
Due to the cylindrical symmetry of the problem, the exciton wave functions
can be labeled by the total orbital angular momentum $L=l_{cm}+l_{rel}$. To
obtain the coefficients $A_{\lambda ,\lambda ^{\prime }}$, the total
Hamiltonian is diagonalized in the space spanned by the product states $\psi
_\lambda ^{cm}\left( {\bf \vec{R}}\right) \psi _{\lambda ^{\prime
}}^{rel}\left( {\bf \vec{r}}\right) $. In the present calculations, we first
solve the single particle problem of center-of-mass and relative
Hamiltonians $H_{cm}$ and $H_{rel}$, keep several hundreds of the single
particle states, and then pick up the low-lying energy levels to construct
several thousands of product states. Note that our numerical diagonalization
scheme is very efficient and essentially exact in the sense that the
accuracy can be improved as required by increasing the total number of
selected product states.

Once the coefficients $A_{\lambda ,\lambda ^{\prime }}$ are obtained, one
can calculate directly the measurable properties, such as the linear optical
susceptibility of the nano-rings, whose imaginary part is related to the
absorption intensity measured by optical emission experiments. In theory,
the linear optical susceptibility is proportional to the dipole matrix
elements between one electron-hole pair $m$ state and the vacuum state,
which in turn is proportional to the oscillator strengths $F_m$. In the
dipole approximation, it is given by \cite{que,song95,bryant} 
\begin{eqnarray}
F_m &=&\left| \int \int d{\bf \vec{R}}d{\bf \vec{r}}\Psi \left( {\bf \vec{R},%
\vec{r}}\right) \delta \left( {\bf \vec{r}}\right) \right| ^2  \nonumber \\
&=&\left| \sum\limits_{\lambda ,\lambda ^{\prime }}A_{\lambda ,\lambda
^{\prime }}\psi _{\lambda ^{\prime }}^{rel}\left( {\bf 0}\right) \int d{\bf 
\vec{R}}\psi _\lambda ^{cm}\left( {\bf \vec{R}}\right) \right| ^2,
\end{eqnarray}
where the factor $\psi _{\lambda ^{\prime }}^{rel}\left( {\bf 0}\right) $
and the integral over ${\bf \vec{R}}$ ensure that only the excitons with $L=0
$ are created by absorbing photons. Therefore, the frequency dependence of
the linear optical susceptibility $\chi \left( \omega \right) $ can be
expressed as \cite{que,song95,bryant} 
\begin{equation}
\chi \left( \omega \right) \propto \sum_m\frac{F_m}{\hbar \omega
-E_g-E_m-i\Gamma },
\end{equation}
where $E_g$ and $E_m$ are the respective semiconducting band gap of InGaAs
and energy levels of the exciton, and $\Gamma $ has been introduced as a
phenomenological broadening parameter.

In what follows we constraint ourselves in the subspace $L=0$, the most
interesting case, throughout the whole calculations. As an interesting
example of a typical system, we have taken the parameters $m_e^{*}=0.067m_e$%
, the effective mass of the heavy hole $m_h^{*}=0.335m_e$ ($m_e$ is the bare
mass of single electron) and $\varepsilon _r=12.4,$ which are appropriate to
InGaAs material\cite{lorke00,emperador,hu}. The electron and hole are
considered to be confined under the same potential barrier, {\em i.e.}, $%
m_e^{*}\omega _e^2=m_h^{*}\omega _h^2$. If we choose the characteristic
energy and length scale to be the effective Rydberg $R_y^{*}=\frac{m_e^{*}e^4%
}{2\hbar ^2\left( 4\pi \varepsilon _0\varepsilon _r\right) ^2}$ and the
effective Bohr radius $a_B^{*}=\frac{4\pi \varepsilon _0\varepsilon _r\hbar
^2}{\mu e^2}$, we find that $R_y^{*}=5.0$ {\rm meV} and $a_B^{*}=11.8$ {\rm %
nm}. In the following, we perform the calculations for two ring radii: $10$
and $30$ {\rm nm}. The ring width can be tuned by the confining potential, 
{\em e.g.}, $W=10$ {\rm nm} corresponds to $\hbar \omega _e=15$ {\rm meV}.

Fig. 2 displays the exciton binding energies obtained for different ring
radii: $R_0=10$ {\rm nm }and{\rm \ }$30${\rm \ nm}, as a function of
nano-ring width.{\rm \ }For comparison, the binding energy of a quantum dot
with a parabolic potential $U\left( {\bf \vec{r}}\right) =\frac 12%
m^{*}\omega _0^2r^2$ is also presented in dashed line (For quantum dots, $%
2W=2\sqrt{\frac \hbar {m^{*}\omega _0}}$ is the diameter). Notice that $%
E_b=E_{e-h}^0-E_{grnd}^{ex}$, where the first term refers to only the
confinement ground state of the electron and hole, ignoring the Coulomb
interaction. It is obvious that for relatively large widths, the exciton
binding energy for small radius nano-rings is larger than for the large
radius ones, as expected. This difference is a reflection of the strong
quantum confinement in small nano-rings. As the ring width decreases, the
binding energy for nano-rings with a large radius increases rapidly. For
widths less than $\approx 11$ {\rm nm}, however, the two solid curves cross
and theirs sequence is reversed. This crossover is caused by the strong
anisotropic confinement in nano-rings: For smaller ring widths, the
resulting exciton wavefunctions are increasingly elongated along the ring,
and thus the exciton is confined in a quasi one-dimensional system with a
characteristic size $\approx W$. On the other hand, by decreasing the ring
radius in a fixed potential strength (or a fixed ring width), nano-rings can
be tuned from quasi one-dimensional to two-dimensional systems. In other
words, nano-rings would behave like quantum dots when theirs radii are
comparable to widths (see Fig. (1a).). One thus can expect that with a fixed
sufficient small ring width, the effective size of the exciton might be
smaller for a larger ring radius, and in turn causes the enhancement of its
binding energy.

Another feature shown in Fig. 2 is the similarity of the curves for the
large radius nano-ring and the quantum dot. This is due to the comparable
confinement area of the two systems, as pointed out by Song and Ulloa that
the excitons in nano-rings behave to a great extent as those in quantum dots
of similar dimensions \cite{song00}. It is also important to emphasize that
for a nano-ring with a large ring radius and ring width, the binding energy
approaches approximately the exact result of a free two-dimensional exciton, 
{\em i.e.}, $E_b=4R_y^{*}=20$ {\rm meV.}

Fig. 3 shows the exciton oscillator strengths versus the ring width, for two
ring radii (two solid lines) and a quantum dot with a parabolic potential
(the dashed line). It is readily seen that the oscillator strength of the
large radius nano-ring is much larger than that of the small radius
nano-ring in the whole range of widths shown, which is also an indication of
the strong quantum confinement in small nano-rings, as mentioned above. For
a larger ring width, the oscillator strengths of the two nano-rings clearly
increase, but not as fast as that of the quantum dot.

To support the experimental relevance of our results, we have also
calculated the linear optical susceptibility of nano-rings. Figs.(4a) and
(4b) show the typical results for different values of the ring width and two
ring radii, where a broadening parameter $\Gamma =0.5$ {\rm meV} is used.
Those curves represent all the possible transitions of excitonic states
which would be measurable via photoluminescence excitation measurements
(PLE). In contrast to the conventional quantum dots, in which the low-lying
exciton state transitions have the same amplitudes and are nearly equally
distributed (a reflection of excitations of the center-of-mass degree of
freedom), the low-lying transitions of nano-rings show a rapid dampen with
frequency and theirs positions are not periodic. This difference is a
refection of the anisotropic confinement of nano-rings: Since the exciton is
confined in a quasi one-dimensional system, its center-of-mass degree of
freedom is greatly suppressed and its relative motion becomes dominant, thus
resulting the destruction of the regular patterns observed in quantum dots.
Note that this behavior is indeed observed in a recent experiment \cite
{pettersson}. Another noticeable feature in Figs. (4a) and (4b) is that
those transition peaks are strongly red shifted as the ring width increases,
indicating the less confinement for larger ring widths.

In conclusion, we have shown the strong quantum confinement effects on
excitons in a nano-ring based on a simple model Hamiltonian. By numerical
diagonalization, we calculate the binding energies, oscillator strengths and
their dependence on the width of nano-rings, as well as the linear optical
susceptibility. The anisotropic confinement in nano-rings is clearly
demonstrated, which can be confirmed by future measurements of optical
emission on InGaAs nano-rings with tunable sizes.

{\bf Acknowledgments}

The financial support from NSF-China (Grant No. 19974019) and China's
''973'' program is gratefully acknowledged.

\begin{center}
{\bf Figures Captions}
\end{center}

Fig.1. The confining potential $U\left( {\bf \vec{r}}\right) =\frac{%
m_e^{*}\omega _e^2}{2R_0^2}\left( {\bf \vec{r}}{}^2-R_0^2\right) ^2$ with
different ring radii $R_0$: $10$ {\rm nm} (a) and $30$ {\rm nm} (b){\rm . }%
The solid and dashed lines correspond to the ring width $W=16.7$ and $9.6$ 
{\rm nm}, respectively.\newline

Fig.2. The exciton binding energies for the nano-ring as a function of ring
width with two different ring radii $R_0=10$ and $30$ {\rm nm}. For
comparison, the result for a parabolic quantum dot is also displayed in
dashed line.\newline

Fig.3. The exciton oscillator strengths versus the ring width, for two ring
radii (two solid lines) and a quantum dot with a parabolic potential (the
dashed line).\newline

Fig.4. Imaginary part of linear optical susceptibility as a function of
frequency $\omega $ for different values of the ring width and two different
ring radii $R_0=10$ {\rm nm} (a) and $R_0=30$ {\rm nm }(b). In each panel,
from bottom to top, the ring widths are $10$, $15$, $20$ and $25$ {\rm nm}.
For clarity, the semiconducting band gap $E_g$ is set to be zero.

\end{document}